\documentclass[aps,physrev,reprint,superscriptaddress]{revtex4-2}
\usepackage{amsmath}
\usepackage{amssymb}
\usepackage{times}
\usepackage{newtxmath}
\usepackage{graphicx} 
\usepackage{xcolor}  
\usepackage{hyperref}
\hypersetup{
    colorlinks=true,    
    linkcolor=blue,     
    filecolor=blue,     
    urlcolor=blue,      
    citecolor=blue      
}

\begin{document}

\title{Ultrahigh free-electron Kerr nonlinearity in all-semiconductor waveguides for all-optical nonlinear modulation of mid-infrared light}

\author{Gonzalo \'Alvarez-P\'erez}
\email{gonzalo.alvarezperez@iit.it}
\affiliation{Istituto Italiano di Tecnologia, Center for Biomolecular Nanotechnologies, Via Barsanti 14, 73010 Arnesano, Italy}
\author{Huatian Hu}
\email{huatian.hu@iit.it}
\affiliation{Istituto Italiano di Tecnologia, Center for Biomolecular Nanotechnologies, Via Barsanti 14, 73010 Arnesano, Italy}
\author{Fangcheng Huang}
\affiliation{Istituto Italiano di Tecnologia, Center for Biomolecular Nanotechnologies, Via Barsanti 14, 73010 Arnesano, Italy}
\author{Tadele Orbula Otomalo}
\affiliation{Istituto Italiano di Tecnologia, Center for Biomolecular Nanotechnologies, Via Barsanti 14, 73010 Arnesano, Italy}
\author{Michele Ortolani}
\affiliation{Istituto di Fotonica e Nanotecnologie, Consiglio Nazionale delle Ricerche, Via del Fosso del Cavaliere 100, Rome, 00133, Italy}
\affiliation{Department of Physics, Sapienza University of Rome, Piazzale Aldo Moro 5, 00185 Rome, Italy}
\affiliation{Istituto Italiano di Tecnologia, Center for Life Nano- and Neuro-Science, Viale Regina Elena 291, Rome, 00161, Italy}
\author{Cristian Cirac\`i}
\email{cristian.ciraci@iit.it}
\affiliation{Istituto Italiano di Tecnologia, Center for Biomolecular Nanotechnologies, Via Barsanti 14, 73010 Arnesano, Italy}

\date{\today}

\begin{abstract}
Nonlinear optical waveguides, particularly those harnessing the optical Kerr effect, are promising for advancing next-generation photonic technologies. Despite the Kerr effect’s ultrafast response, its inherently weak nonlinearity has hindered practical applications. Here, we explore free-electron-induced Kerr nonlinearities in all-semiconductor waveguides, revealing that longitudinal bulk plasmons—inherently nonlocal excitations—can generate exceptionally strong Kerr nonlinearities. We specifically develop a nonlinear eigenmode analysis integrated with semiclassical hydrodynamic theory to compute the linear and nonlinear optical responses originating from the quantum behavior of free electrons in heavily doped semiconductors. These waveguides achieve ultrahigh nonlinear coefficients exceeding 10\textsuperscript{7} W\textsuperscript{-1}km\textsuperscript{-1} and support long-propagating modes with propagation distances over 100 $\mu$m. Additionally, we confirm the robustness of the nonlinear response under realistic conditions by considering viscoelastic and nonlinear damping mechanisms. Finally, we implement our all-semiconductor waveguides in a Mach-Zehnder interferometer, demonstrating efficient nonlinear modulation of the transmittance spectrum via the free-electron Kerr effect. This work evidences the transformative potential of free-electron nonlinearities in heavily doped semiconductors for photonic integrated circuits, paving the way for scalable on-chip nonlinear nanophotonic systems.
\end{abstract}


\maketitle

\section{Introduction}

Nonlinear optical waveguides are at the core of modern photonic technologies and are increasingly promising for next-generation systems in telecommunications, quantum technologies, optical computing, and biomedical sensing \cite{suhara2003waveguide}. Among these, Kerr waveguides leverage the optical Kerr effect—a nonlinear phenomenon where the refractive index changes with light intensity \cite{boyd2008nonlinear}—to enable critical processes such as self-phase modulation and soliton formation. In fact, refractive index engineering is central to many advancements in both traditional and meta optics \cite{smith2004metamaterials,galiffi2022photonics}. By harnessing the optical Kerr effect, it is possible to achieve efficient and ultrafast photon–photon interactions, which are essential for applications like high-speed optical communication, wavelength conversion, entangled-photon generation, optical switching, modulation, and frequency-comb formation \cite{krasavin2018free,almeida2004optical,gibbs1982room,un2023electronic,cox2014electrically,christensen2015kerr,cox2016quantum,ryabov2022nonlinear,niu2023all}. Additionally, the growing demands of artificial intelligence platforms have reignited interest in Kerr-based solutions, particularly due to their potential to deliver integrated, low-power, and ultrafast optical modulation. However, despite the Kerr effect's inherent speed—operating on femtosecond timescales—its relatively weak nonlinearity poses a challenge. Conversely, thermal nonlinearities, while potentially stronger, are limited by their slow response times (kHz–MHz timescales) \cite{almeida2004optical}. Achieving both modulation strength and speed remains therefore a critical challenge for advancing photonic technologies.

Kerr guided-wave optics has long relied on silica fibers, which possess sub-10-$\mu$m core diameters and ultralow propagation losses. To characterize the Kerr nonlinearity in waveguides, which we denote here by $\gamma_\mathrm{wg}$, it is conventional to use the ratio between the nonlinear refractive index and the effective mode area. Silica fibers have a relatively small nonlinear coefficient ($\gamma_\mathrm{wg} \approx 20$ W\textsuperscript{-1}km\textsuperscript{-1}), typically requiring extended (meter- to kilometer-scale) interaction lengths \cite{Toulouse2005}, which makes them suboptimal despite their low losses. Photonic crystal fibers partly address these constraints by using high-nonlinearity SF57 glass and tighter mode confinement, achieving coefficients up to 640~W\textsuperscript{-1}km\textsuperscript{-1} \cite{Petropoulos2003}. Silicon-on-insulator (SOI) Kerr on-chip integrated waveguides further enhance nonlinear response through silicon's large Kerr nonlinearity and strong modal confinement, reaching $\gamma_\mathrm{wg} \approx 10^5$~W\textsuperscript{-1}km\textsuperscript{-1} \cite{Koos2007,Leuthold2010}, but two-photon absorption and free-carrier losses at telecom wavelengths limit both their speed and functionality \cite{Moss2013,Lin2007}. Silicon--organic slot waveguides mitigate these issues by confining light within highly nonlinear polymers, attaining $\gamma_\mathrm{wg} \sim 10^6$~W\textsuperscript{-1}km\textsuperscript{-1} \cite{Koos2007,Koos2009}. Nevertheless, diffraction commonly constrains traditional Kerr guided-wave optics lengths to a few millimeters. Plasmonic platforms can overcome these restrictions by localizing electromagnetic fields below the diffraction limit, enhancing light--matter interactions and enabling the miniaturization of Kerr devices to the nanometer scale \cite{Ebbesen2008,Kauranen2012,Salgueiro2012,Li2016_figure,Li2018_fundamentals}. Theoretical projections suggest nonlinear coefficients on the order of $10^4$~W\textsuperscript{-1}km\textsuperscript{-1} for nanoshell plasmonic waveguides \cite{Hossain2011}, $10^7$~W\textsuperscript{-1}km\textsuperscript{-1} in metal-indium tin oxide (ITO)-metal slot waveguides \cite{RojasYanez25}, while four-wave mixing in short hybrid-plasmonic devices \cite{Diaz2016} holds promise for improved performance in ultra-compact geometries. However, these systems often suffer from high losses or limited scalability.

In this context, heavily doped semiconductors have emerged as promising alternative plasmonic materials for the infrared band. Apart from benefiting from mature large-scale fabrication, they can strongly confine electromagnetic fields while supporting strong, tunable free-electron (FE) optical nonlinearities \cite{lee2014giant,yu2022electrically}. Doped semiconductors transition from the size-quantization regime of quantum dots and wells to the classical regime of plasmon oscillations, where electron-electron interactions can induce strong nonlocal and nonlinear optical responses \cite{de2021free}. These collective FE oscillations can surpass traditional metal nonlinearities, partly because lower electron densities ($10^{18}$--$10^{19}$~cm\textsuperscript{-3}) enhance the third-order polarizability, which, in the limit of small excitations, scales inversely with the square of the FE density, $n_0^2$ \cite{rossetti2024origin}. Doped InP and InGaAs stand out as promising candidates for exhibiting such high FE nonlinearities \cite{de2022material}, with third-harmonic generation exceeding conventional $\chi^{(3)}$ nonlinearities, as recently observed in doped InGaAs nanoantennas \cite{rossetti2024origin}. Additionally, surface carrier density modulation enables reconfigurable FE nonlinearities on ultrafast timescales, outpacing slower thermal mechanisms \cite{de2022impact,almeida2004optical,hu2024modulating}.

Despite their promise, FE-driven effects in heavily doped semiconductors have seen limited deployment in waveguides and on-chip integrated circuits. Past research has largely emphasized bulk $\chi^{(3)}$ processes over the many-micrometer propagation distances typical of waveguides. Although second-harmonic generation via FEs has been reported in metal-insulator-metal waveguides \cite{noor22}, FE-driven nonlinearities in semiconductor waveguides---in particular the Kerr effect---have remained elusive. One reason is the significant scale mismatch between the highly localized volumes where FE responses dominate---typically on the order of a few nanometers---and the long propagation lengths of guided modes in hybrid plasmonic waveguides---from hundreds to thousands of microns---, which impose stringent computational demands for their design and optimization. Another reason is that FEs typically respond below the plasma frequency, restricting bulk Kerr effects. However, recent work \cite{hu2024low} has shown that heavily doped semiconductors can support strong nanoscale Kerr nonlinearities above the plasma frequency via longitudinal bulk plasmons (LBPs)---intrinsically nonlocal excitations fulfilling $\varepsilon(\omega,\mathbf{k})=0$ (with $\omega$ and $\mathbf{k}$ being frequency and wavevector, respectively) \cite{hu2024low,ruppin_optical_1973,ruppin2001extinction,raza2011unusual}. Historically identified in thin metal films at ultraviolet energies \cite{anderegg_optically_1971,lindau_experimental_1971}, LBPs have recently been observed at infrared frequencies in transparent conducting oxides \cite{de2018viscoelastic} and n-doped InAsSb \cite{moireau2024bulk}. In this work, we integrate LBPs into hybrid designs featuring intrinsic waveguide cores made of undoped III-V semiconductors evanescently coupled to heavily doped III-V semiconductor layers. Using a nonlinear hydrodynamic eigenmode analysis method that we specifically develop here, we demonstrate that the FE Kerr effect enhances the waveguide nonlinear coefficient in these hybrid structures to \(\gamma_{\mathrm{vg}} \approx 4 \times 10^7\,\mathrm{W}^{-1}\,\mathrm{km}^{-1}\). Crucially, this boost occurs without compromising the low-loss properties inherent to undoped III-V semiconductors, enabling strong nonlinearities over propagation distances exceeding 100 $\mu$m. We also demonstrate nonlinear modulation of the transmission spectrum in a Mach-Zehnder interferometer driven by the FE Kerr effect, introducing an ultrafast photonic platform for telecom to mid-infrared frequencies.

\section{Hydrodynamic linear response of hybrid heavily doped semiconductor waveguides}

\begin{figure*}[!ht]
\includegraphics[width=\textwidth]{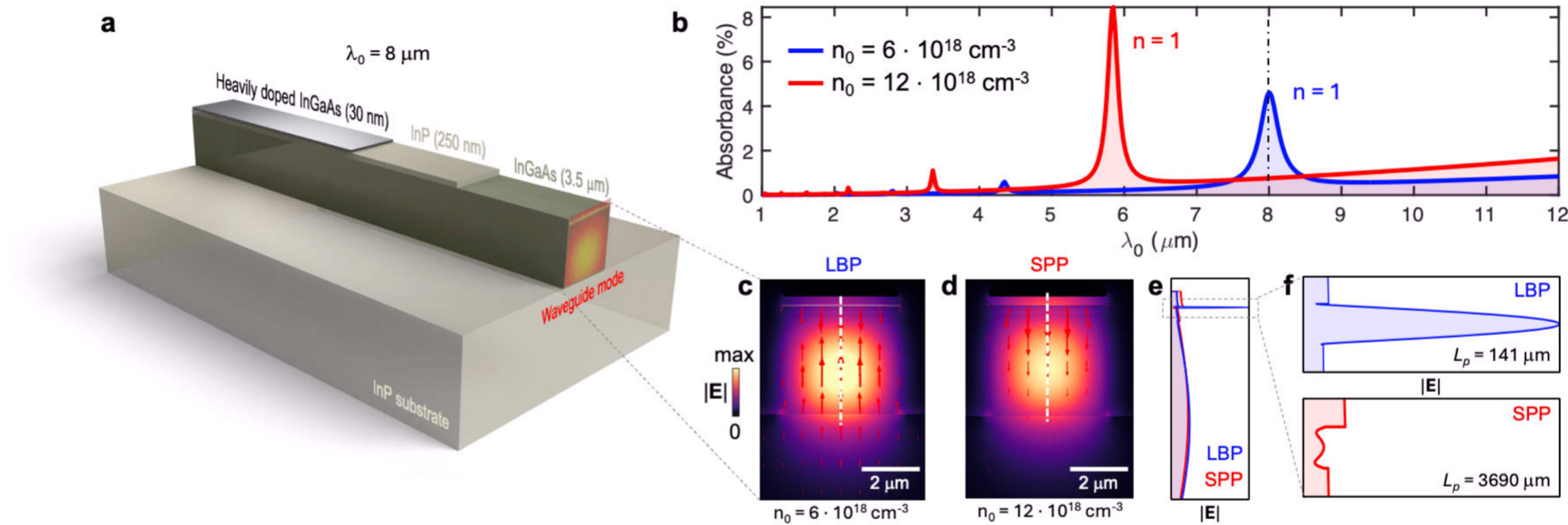}
\caption{\textbf{Hydrodynamic linear optical response of hybrid all-semiconductor waveguides. (a)} Schematic of the structure, consisting of a 3.5 $\mu$m thick InGaAs core and an InP cladding, which supports a guided mode. On top, a 30-nm thick slab of heavily doped InGaAs is placed. \textbf{(b)} Absorbance spectra of the heavily doped InGaAs slab for the two different doping concentrations: $n_0 = 6 \cdot 10^{18}$ cm\textsuperscript{-3} (blue curve) and $n_0 = 12 \cdot 10^{18}$ cm\textsuperscript{-3} (red curve). The dashed line marks $\lambda_0=8$ $\mu$m. \textbf{(c, d)} Simulated TM (transverse magnetic) electric field distributions for $n_0 = 6 \cdot 10^{18}$ cm\textsuperscript{-3} (LBP mode) and $n_0 = 12 \cdot 10^{18}$ cm\textsuperscript{-3} (SPP mode) at $\lambda_0=8$ $\mu$m. \textbf{(e)} Line profile of the normalized electric field amplitude $|\textbf{E}|$ along the center of the structure for both doping levels. The LBP mode (blue curve) shows a significant field amplitude compared to the SPP mode (red curve). \textbf{(f)} Zoomed-in views of the electric field profiles for the LBP and SPP modes. The LBP mode exhibits a propagation length $L_p=141\,\mu$m due to strong interaction between the guided mode and the heavily doped InGaAs layer, while the SPP mode has a much longer propagation length of $L_p=3690\,\mu$m due to lower interaction.}
\label{fig:1}
\end{figure*}

From a theoretical perspective, LBP resonances can be excited in finite-size systems if the material response is nonlocal \cite{ruppin2001extinction,christensen2014nonlocal}. One way to account for nonlocality in degenerate electron systems is to model the classical FE dynamics together with density-dependent energy potentials that encapsulate their quantum properties. This approach is commonly referred to as hydrodynamic theory (HT) \cite{ciraci2016quantum, toscano2015resonance}, deriving its name from the fact that FEs are intuitively treated as a fluid. 
In a preliminary work \cite{hu2024low}, we developed a time-domain electromagnetic model integrated with semiclassical HT that directly captures the linear and nonlinear optical responses arising from the quantum nature of electrons—an often nontrivial task for conventional semiconductor frameworks based on the Schrödinger-Poisson equations \cite{silberberg1992optical, qian2016giant, cominotti2023theory}. Accounting for plasmonic hydrodynamic contributions to Kerr nonlinearity in heavily doped semiconductors, we discovered a strong FE Kerr-type nonlinearity that enables low-threshold bistability—two stable outputs at the same input intensity of 1 mW. Here, we develop a nonlinear hydrodynamic eigenmode analysis and extend this model to heavily doped semiconductor waveguides, showing that LBPs can indeed drive a very large Kerr effect with strong modulation of the mode's effective refractive index.

Our proposed nonlinear all-semiconductor system and its linear optical response are described in Fig. \ref{fig:1}. It consists of a 3.5\,$\mu$m-thick undoped InGaAs core between an InP substrate and a 250-nm-thick InP layer, with a 30-nm-thick heavily doped InGaAs slab on top to modulate the guided mode (Fig. \ref{fig:1}a). This configuration supports both transverse electric (TE) and transverse magnetic (TM) modes (see Supplementary Section S1). To decode the optical response of the system, it is key to understand the motion of FEs in the heavily doped InGaAs layer. Following the HT \cite{de2021free,de2022impact,hu2024low}, the equation of motion involves two macroscopic quantities: the electron density \(n(\mathbf{r},t)\) and velocity \(\mathbf{v}(\mathbf{r},t)\):
\begin{equation}\label{eq:HT-EqMotion}
m_e\!\Bigl(\frac{\partial}{\partial t} + \mathbf{v} \cdot \nabla + \gamma\Bigr)\mathbf{v} 
= -e\bigl(\mathbf{E} + \mu_0 \mathbf{v} \times \mathbf{H}\bigr) 
- \nabla \frac{\delta G[n]}{\delta n},
\end{equation}
where \(\mathbf{E}\) and \(\mathbf{H}\) are the electric and magnetic fields, respectively, and \(m_e\), \(e\), and \(\mu_0\) are the effective electron mass, electron charge, and vacuum permeability, respectively. Equation \eqref{eq:HT-EqMotion} incorporates convection (\(\mathbf{v}\!\cdot\!\nabla\mathbf{v}\)), dissipation (\(\gamma\)), Coulomb and Lorentz forces, and a quantum pressure term arising from the internal energy functional \(G[n]\). The total electron density is \(n=n_0(\mathbf{r}) + \sum_{j}n_j(t,\mathbf{r})\), where \(n_0(\mathbf{r})\) is the equilibrium density and \(n_j\) the perturbed densities at harmonic order \(j\). We adopt the Thomas-Fermi (TF) approximation for \(G[n]\simeq T_{\mathrm{TF}}[n]\), which is crucial for describing nonlocal effects \cite{ruppin2001extinction}. Additional refinements, including electron spill-out, can shift the plasmon resonance slightly \cite{toscano2015resonance,ciraci2016quantum}, but these are neglected here since they are typically relevant only in single-digit nanometer scales and below. By rewriting Eq.~\eqref{eq:HT-EqMotion} with \(\dot{\mathbf{P}}=\mathbf{J}=-ne\mathbf{v}\), we express the optical response through the constitutive relation at the \(j^\text{th}\) harmonic \cite{de2021free}:
\begin{equation}\label{eq:constitutive}
\ddot{\mathbf{P}}_j + \gamma \dot{\mathbf{P}}_j 
= \frac{n_0 e^2}{m_e}\,\mathbf{E}_j 
+ \beta^2 \nabla (\nabla \!\cdot\! \mathbf{P}_j) 
+ \mathbf{S}^{\mathrm{NL}}_{\omega_j},
\end{equation}
where the first-order TF quantum-pressure term is \(\tfrac{en_0}{m_e}\nabla\bigl(\tfrac{\delta T_\mathrm{TF}}{\delta n}\bigr)_1 \!=\! \beta^2 \nabla (\nabla \!\cdot\! \mathbf{P}_j)\). Here, \(\beta^2=\tfrac{3}{5}v_{\rm F}^2 = 2\tfrac{c_{\rm TF}}{m_e}n_0^{2/3}\), \(c_{\mathrm{TF}}=\tfrac{\hbar}{m_e}\tfrac{3}{10}(3\pi^2)^{2/3}\), where \(v_{\mathrm{F}}\) is the Fermi velocity \cite{grosso2013solid}. The ``$\cdot$" on the variables denotes the time derivative and  \(\mathbf{S}^{\mathrm{NL}}_{\omega_j}\) is the FE nonlinear source. By combining Eq. \eqref{eq:constitutive} and Maxwell’s equations in the frequency domain---\textit{i.e.}, assuming time-harmonic fields \(\mathbf{F}(\mathbf{r},t) = \sum_j \mathbf{F}_j(\mathbf{r}) e^{-i\omega_j t}\), where \(\mathbf{F} \in \{\mathbf{E}, \mathbf{H}, \mathbf{P}\}\)---we have:
\begin{align}
    \label{eq:freq_domain}
    &\nabla \times \nabla \times \mathbf{E}_j 
    - \varepsilon\frac{\omega_j^2}{c^2} \mathbf{E}_j 
    - \mu_0\omega_j^2 (\mathbf{P}^{\text{NL}}_{\omega_j}+\mathbf{P}_j) = 0,\\
    \label{eq:freq_domain_P}
    &\beta^2 \nabla (\nabla \cdot \mathbf{P}_j) 
    + (\omega_j^2 + i \gamma \omega_j) \mathbf{P}_j = -\frac{n_0 e^2}{m_e} \mathbf{E}_j + \mathbf{S}^{\mathrm{NL}}_{\omega_j}.
\end{align}

In Eqs. (\ref{eq:freq_domain},\ref{eq:freq_domain_P}), we account for the dielectric local contributions from the semiconductor, including both linear effects, represented by the local permittivity $\varepsilon$, and nonlinear effects, characterized by the nonlinear polarization $\mathbf{P}^{\text{NL}}_{\omega_j}$. The coupling between different harmonic frequencies arises from the nonlinear contributions $\mathbf{P}^{\text{NL}}_{\omega_j}$ and $\mathbf{S}^{\mathrm{NL}}_{\omega_j}$. For simplicity, we assume that the pump field remains unaffected by the nonlinear process (undepleted pump approximation), i.e., $\mathbf{P}^{\text{NL}}_{\omega_1} = \mathbf{S}^\text{NL}_{\omega_1} = 0$, as harmonic signals are expected to be several orders of magnitude weaker than the pump field. The crystal lattice nonlinearities are incorporated through a bulk third-order susceptibility $\chi^{(3)}$, described by $\mathbf{P}^{\text{NL}}_{\omega_3} = \varepsilon_0 \chi^{(3)} (\mathbf{E}_1 \cdot \mathbf{E}_1)\mathbf{E}_1$. Moreover, we impose the continuity of the normal polarization component, \(P_n^- = P_n^+\) as the required additional boundary condition \cite{scalora2010second,ciraci2012second,krasavin2018free,Raza2013waveguides,Huang13,Toscano2013waveguiding}, and take \(n_0\) constant in the metal, zero outside (hard-wall condition).

We first analyze the linear optical response (\(j=1\)) from Eqs. (\ref{eq:freq_domain},\ref{eq:freq_domain_P}) (with \(\mathbf{S}^{\mathrm{NL}}_{\omega_j}\!=\!0\)). We consider an infinite, 30-nm-thick slab of heavily doped InGaAs with carrier density \(n_0\) under a plane wave with normal incidence. The material follows a Drude-like response with \(\varepsilon_\infty\!=\!12\), \(\gamma=8.9\,\mathrm{ps}^{-1}\), and \(m_e=0.041m_0\) (\(m_0\) is the electron mass) \cite{rossetti2024origin,hu2024low}. The screened bulk plasma wavelength is \(\lambda_p\!=\!2\pi c\,\sqrt{m_e\varepsilon_0\varepsilon_\infty/(n_0e^2)}\). Figure \ref{fig:1}b shows the slab absorbance for two doping levels: \(n_0 = 12 \times 10^{18}\,\mathrm{cm}^{-3}\) (red) and \(n_0 = 6 \times 10^{18}\,\mathrm{cm}^{-3}\) (blue),  calculated using a finite-element eigenmode solver (we use \textsc{Comsol Multiphysics}) \cite{Huang13,Toscano2013waveguiding,Zheng2019}. The bulk plasmon or epsilon-near-zero (ENZ) resonances for both systems are at \(\lambda_0=5.85\,\mu\mathrm{m}\) and \(\lambda_0=8\,\mu\mathrm{m}\), respectively, with additional higher-order resonances satisfying \(\varepsilon(\omega,\mathbf{k})=0\) under HT. Only odd-order LBPs carry a net dipole and thus couple to normal incidence \cite{hu2024low}. At \(\lambda_0=8\,\mu\mathrm{m}\), the slab with \(n_0=12 \times 10^{18}\,\mathrm{cm}^{-3}\) (with screened plasma wavelength at 7.33 $\mu$m) supports an SPP, with LBP resonances occurring at lower wavelengths, whereas for a doping level \(n_0=6 \times 10^{18}\,\mathrm{cm}^{-3}\) (with screened plasma wavelength at 10.37 $\mu$m) \(\lambda_0=8\,\mu\mathrm{m}\) exactly matches the first ($n=1$) LBP mode.

We can then find out the linear modes supported by the structure shown in Fig. \ref{fig:1}a by solving the corresponding eigenvalue problem. From Eq. \eqref{eq:freq_domain_P}, and using the curl of curl identity, $\nabla \times (\nabla \times \mathbf{E}_j) = \nabla(\nabla \cdot \mathbf{E}_j) - \nabla^2 \mathbf{E}_j
$, together with \(\nabla \!\cdot \!\mathbf{P}_j = -\varepsilon_0 \nabla \!\cdot\! \mathbf{E}_j\) and rearranging terms, we have:
\begin{equation}
    \mathbf{P}_j = \varepsilon_0 \,\chi_j \!\left[\mathbf{E}_j 
    - \frac{\beta^2}{\omega_p^2}\,\nabla \nabla \!\cdot \!\mathbf{E}_j\right] 
    + \frac{\chi_j}{\omega_p^2} \mathbf{S}^{\mathrm{NL}}_{\omega_j},
    \label{eq:polarization}
\end{equation}
where \(\chi_j = \varepsilon(\omega_j) - 1 
= -\tfrac{\omega_p^2}{\omega_j^2 + i \gamma \omega_j}\). Then, substituting Eq. \eqref{eq:polarization} into Eq.~\eqref{eq:freq_domain}, using again the curl of curl identity, and particularizing for $j=1$ yields the eigenvalue problem for the fundamental mode, where the eigenvalue is the wavevector $k_1$:
\begin{align}\label{eq:eigenvalue}
\nabla^2 \mathbf{E}_1 
- \biggl[1 - \frac{\beta^2 k_1^2 \,\chi(\omega)}{\omega_p^2}\biggr] 
\nabla \nabla \!\cdot \!\mathbf{E}_1 
+ \varepsilon(\omega)\,k_1^2 \,\mathbf{E}_1 = 0,
\end{align}
where we have rewritten $k_1^2=\varepsilon\omega^2/c^2$ and 
assumed \(\mathbf{S}_{\omega_1}^{\mathrm{NL}} \approx 0\) (undepleted pump approximation). Letting the mode propagate along \(z\), \(\mathbf{E}_1(\mathbf{r}) = A_1\,\tilde{\mathbf{E}}_1(x,y)\,e^{i\kappa_1 z}\)
. We solve Eq.~\eqref{eq:eigenvalue} numerically for arbitrary cross-sections using a finite-element eigenmode solver (we use \textsc{Comsol Multiphysics}) \cite{Huang13,Toscano2013waveguiding,Zheng2019}.

Figures \ref{fig:1}c,d show the normalized electric-field amplitudes \(|\mathbf{E}|\) of the TM mode for both doping levels, \(n_0 = 6 \times 10^{18}\,\mathrm{cm}^{-3}\) (Fig. \ref{fig:1}c) and \(n_0 = 12 \times 10^{18}\,\mathrm{cm}^{-3}\) (Fig. \ref{fig:1}d) at $z=0$. In each case, the mode is mostly confined in the undoped InGaAs and extends into the heavily doped layer. The field profiles in Fig. \ref{fig:1}e reveal a stark difference between both cases: the LBP mode displays a significantly larger field amplitude in the bulk of doped layer, while the SPP mode’s field amplitude is comparatively smaller and confined in the vicinity of the surface. More importantly, the former mode has the maximum field amplitude inside the heavily doped InGaAs whereas the latter repels the field outside of the InGaAs, where the nonlinearity comes from. A closer view (Fig. \ref{fig:1}f) underscores this difference and indicates the respective propagation lengths. The LBP mode has \(L_p = 141\,\mu\mathrm{m}\), enabled by strong interaction between the guided mode and the doped layer, whereas the SPP mode shows a longer \(L_p=3690\,\mu\mathrm{m}\) due to weaker overlap. Overall, the hybrid LBP mode provides stronger field confinement and enhanced FE interactions, leveraging both the low-loss waveguide mode and the strong contribution of the heavily doped InGaAs.

\section{Free-electron Kerr nonlinearity in hybrid heavily doped semiconductor waveguides}

In fact, the LBP is a charge density wave in the bulk (fields shown as Fig. \ref{fig:1}c) that overlaps with the physical dimension of the heavily doped InGaAs, guaranteeing enough active interaction volume and strong nonlinearity \cite{hu2024low}. This hybrid LBP-guided mode is thus an ideal platform for generating strong FE Kerr at low power, which we now investigate. To calculate the nonlinear modulation due to the FE-driven Kerr effect, we solve the nonlinear problem self-consistently to allow the field to be self-modulated by its own intensity. In particular, we solve the extension of Eq. \eqref{eq:eigenvalue} to Kerr nonlinearity by accounting for third-order (i.e., direct effects) processes, as indicated by $\mathbf{S} ^{(3)}_{\omega _1}$, neglecting cascaded effects for simplicity:

\begin{equation}
    \nabla^2 \tilde{\mathbf{E}}_1' - \left[ 1 - \frac{\beta^2 k_1'^2 \chi(\omega)}{\omega_p^2} \right] \nabla \nabla \cdot \tilde{\mathbf{E}}_1' + \varepsilon(\omega) k_1'^2 \tilde{\mathbf{E}}_1'+\mathbf{S} ^{(3)}_{\omega _1}= 0,
    \label{eq:nonlinear_eigenvalue}
\end{equation}
where $\tilde{\mathbf{E}}_1'$ and $k_1'^2$ are the self-consistent solutions for the transverse electric field and wavevector under FE Kerr nonlinearity, respectively. $(\mathbf{S}^{(3)}_{\omega _1})'$ is the self-consistent FE Kerr source, which is given by:
\begin{widetext} 
\begin{equation}
\label{eq:direct}
\begin{aligned}
(\mathbf{S}^{(3)}_{\omega _1})' & = -\frac{1}{e^2n_0^2} \left[  \nabla \cdot (\mathbf{P}_1')^*\dot{\bf P}_1'\nabla \cdot \dot{\bf P}_1' +  \nabla \cdot \mathbf{P}_1'(\dot{\bf P}')^*_1\nabla \cdot \dot{\bf P}_1' +  \nabla \cdot \mathbf{P}_1'\dot{\bf P}_1'\nabla \cdot (\dot{\mathbf{P}}_1')^*
+\nabla \cdot (\mathbf{P}_1')^*\dot{\bf P}_1'\cdot \nabla \dot{\bf P}_1'+\nabla \cdot \mathbf{P}_1'(\dot{\mathbf{P}}_1')^*\cdot \nabla \dot{\bf P}_1' \right.\\
&\left.+\nabla \cdot \mathbf{P}_1'\dot{\bf P}_1'\cdot \nabla (\dot{\mathbf{P}}_1')^*
+2(\dot{\mathbf{P}}_1')^*\dot{\mathbf{P}}_1'\nabla \nabla \cdot\mathbf{P}_1'+\dot{\mathbf{P}}_1'\dot{\mathbf{P}}_1'\nabla \nabla \cdot(\mathbf{P}_1')^*\right]-\frac{4}{27}\frac{\beta^2}{e^2n_0^3}\left\lbrace\frac{3}{4}n_0\nabla\left[\nabla \cdot (\mathbf{P}_1')^*(\nabla \cdot \mathbf{P}_1)'^2\right]\right\rbrace.
\end{aligned}
\end{equation}
\end{widetext}
with $\mathbf{P}_1'$ denoting the self-consistent solution of the polarization. The ``$*$" on the variables denotes the complex conjugate. Eq. \eqref{eq:direct} contains terms proportional to $n_0^{-2}$, which boost nonlinearities at relatively low doping levels. On the other hand, the Kerr nonlinearity from the dielectric can be considered with a self-consistent polarization $\left(\mathbf{P}^{\mathrm{NL}}_{\mathrm{d}}\right)'=3\varepsilon_0\chi^{(3)}|\tilde{\mathbf{E}}_1'|^2\tilde{\mathbf{E}}_1'$ with the susceptibility of InGaAs given by $\chi^{(3)}_\text{InGaAs}=1.6\times 10^{-18}$ m$^2/$V$^2$, while that of InP is $\chi^{(3)}_\text{InP}=1\times 10^{-18}$ m$^2/$V$^2$ \cite{boyd2008nonlinear}, with \(\varepsilon_0\) the permittivity of free space. 

Let us now consider Eqs. \eqref{eq:nonlinear_eigenvalue} and \eqref{eq:direct}. In nonlinear optics, the divergence term in \eqref{eq:nonlinear_eigenvalue} is generally neglected and a solution can be easily obtained in the slowly varying envelope approximation, through the definition of overlap integrals evaluated in the waveguide cross-section \cite{Ruan2009,Davoyan2010,Zhang2013a,Zhang2013b,Wu2014,Sun2015,Huang2016,Shi2019}. In the case of metal nonlinearities, and in particular of hydrodynamic nonlinearities,
neglecting the divergence will strongly aﬀect the results, since the larger nonlinear contributions arise at the metal
surface, where the divergence is non-zero. On the other hand, fully solving Eqs. \eqref{eq:nonlinear_eigenvalue} and \eqref{eq:direct} in a three-dimensional numerical set-up is challenging, due to the large scale mismatch between the surface eﬀects and the overall mode propagation. Assuming, without loss of generality, that
the modes propagate along the $z$ direction, we can solve for the mode profile at the waveguide cross-section. By writing $\nabla=\nabla_\bot + i\kappa_1\hat{\mathbf{z}}$, the FE Kerr nonlinear source in Eq. \eqref{eq:direct}, it can be rewritten as:

\begin{widetext}
\begin{equation}
\label{eq:direct_z}
\begin{aligned}
(\tilde{\mathbf{S}}^{(3)}_{\omega _1})' = & -\frac{1}{e^2 n_0^2} \bigg[
    \left( \nabla_\bot + i\kappa_1\hat{\mathbf{z}} \right) \cdot (\mathbf{P}_1')^* \dot{\mathbf{P}}_1' \left( \nabla_\bot + i\kappa_1\hat{\mathbf{z}} \right) \cdot \dot{\mathbf{P}}_1' 
    + \left( \nabla_\bot + i\kappa_1\hat{\mathbf{z}} \right) \cdot \mathbf{P}_1' (\dot{\mathbf{P}}_1')^* \left( \nabla_\bot + i\kappa_1\hat{\mathbf{z}} \right) \cdot \dot{\mathbf{P}}_1' \\
    & + \left( \nabla_\bot + i\kappa_1\hat{\mathbf{z}} \right) \cdot \mathbf{P}_1' \dot{\mathbf{P}}_1' \left( \nabla_\bot + i\kappa_1\hat{\mathbf{z}} \right) \cdot (\dot{\mathbf{P}}_1')^* 
    + \left( \nabla_\bot + i\kappa_1\hat{\mathbf{z}} \right) \cdot (\mathbf{P}_1')^* \dot{\mathbf{P}}_1' \cdot \left( \nabla_\bot + i\kappa_1\hat{\mathbf{z}} \right) \dot{\mathbf{P}}_1' \\
    & + \left( \nabla_\bot + i\kappa_1\hat{\mathbf{z}} \right) \cdot \mathbf{P}_1' (\dot{\mathbf{P}}_1')^* \cdot \left( \nabla_\bot + i\kappa_1\hat{\mathbf{z}} \right) \dot{\mathbf{P}}_1' 
    + \left( \nabla_\bot + i\kappa_1\hat{\mathbf{z}} \right) \cdot \mathbf{P}_1' \dot{\mathbf{P}}_1' \cdot \left( \nabla_\bot + i\kappa_1\hat{\mathbf{z}} \right) (\dot{\mathbf{P}}_1')^* \\
    & + 2 (\dot{\mathbf{P}}_1')^* \dot{\mathbf{P}}_1' \left( \nabla_\bot + i\kappa_1\hat{\mathbf{z}} \right)^2 \cdot \mathbf{P}_1' 
    + \dot{\mathbf{P}}_1' \dot{\mathbf{P}}_1' \left( \nabla_\bot + i\kappa_1\hat{\mathbf{z}} \right)^2 \cdot (\mathbf{P}_1')^* 
\bigg] \\
& - \frac{4}{27} \frac{\beta^2}{e^2 n_0^3} \left\lbrace 
    \frac{3}{4} n_0 \left( \nabla_\bot + i\kappa_1\hat{\mathbf{z}} \right) 
    \left[ \left( \nabla_\bot + i\kappa_1\hat{\mathbf{z}} \right) \cdot (\mathbf{P}_1')^* 
    \left( \left( \nabla_\bot + i\kappa_1\hat{\mathbf{z}} \right) \cdot \mathbf{P}_1' \right)^2 
    \right] 
\right\rbrace.
\end{aligned}
\end{equation}
\end{widetext}

We develop a specific nonlinear eigenmode solver to find out the solutions of Eqs. \eqref{eq:nonlinear_eigenvalue} and \eqref{eq:direct_z} and calculate the refractive index change $\Delta n$ in the waveguide due to the FE Kerr effect. In a standard linear mode analysis, one assumes a refractive index profile that is independent of the optical field 
intensity and solves for the corresponding electromagnetic eigenmode. However, in the nonlinear eigenmode problem given by Eqs. \eqref{eq:nonlinear_eigenvalue} and \eqref{eq:direct_z}, the Kerr effect induces a change in the refractive index that depends on the local field intensity. This creates a feedback loop: the mode profile itself modifies the refractive index distribution, which then further modifies the mode. One must then solve Maxwell's equations self-consistently with an intensity-dependent refractive index. Commercially available FEM nonlinear eigenmode solvers can struggle with nonlinearities and, more importantly, with the added complexity of non-localities introduced by functionals like the TF. In our method, we introduce a nonlinear scaling coefficient \(\mathcal{E}_0^2\) in the nonlinear source terms $\tilde{\mathbf{S}}^{(3)}_{\omega _1}$, given by Eq. \eqref{eq:direct_z}, and $\mathbf{P}^{\mathrm{NL}}_{\mathrm{d}}$. We can thus rewrite the eigenvalue problem as:

\begin{equation}
    \begin{aligned}
    &\nabla^2 \tilde{\mathbf{E}}_1' - \left[ 1 - \frac{\beta^2 k_1'^2 \chi(\omega)}{\omega_p^2} \right] \nabla \nabla \cdot \tilde{\mathbf{E}}_1' + \varepsilon(\omega) k_1'^2 \tilde{\mathbf{E}}_1'\\
    &+3\mathcal{E}_0^2\frac{\omega_1^2}{c}\chi^{(3)}|\tilde{\mathbf{E}}_1'|^2\tilde{\mathbf{E}}_1'+\mathcal{E}_0^2\mathbf{S} ^{(3)}_{\omega _1}= 0,
    \label{eq:nonlinear_eigenvalue_final}
    \end{aligned}
\end{equation}
starting from the solution to Eq. \eqref{eq:nonlinear_eigenvalue_final} with $\mathcal{E}_0=0$---i.e., the linear eigenmode---as initial guess for the field amplitude $\tilde{\textbf{E}}_1$ and effective index $n_\text{eff}=k_1/k_0$. Then, we incrementally adjust \(\mathcal{E}_0^2\) in small steps, solving Eq. \eqref{eq:nonlinear_eigenvalue_final} at each step. Each iteration effectively \textit{linearizes} the problem, and we can solve it self-consistently until the effective mode index \(n_\text{eff}'=k_1'/k_0\) converges. Once convergence is reached, $\Delta n=n_\text{eff}'-n_\text{eff}$ and the total input power can be calculated as
$P = \mathcal{E}_0^2 \cdot P_0,$ where $P_0 = \int \mathbf{S}\,\cdot\,\hat{\mathbf{z}}\,dz$ represents the integral of the Poynting vector $z$ component over the waveguide 
cross-section. Because each iteration updates the material response, these converged solutions form the \textit{nonlinear mode}. 

\begin{figure}[ht]
\includegraphics[width=0.35\textwidth]{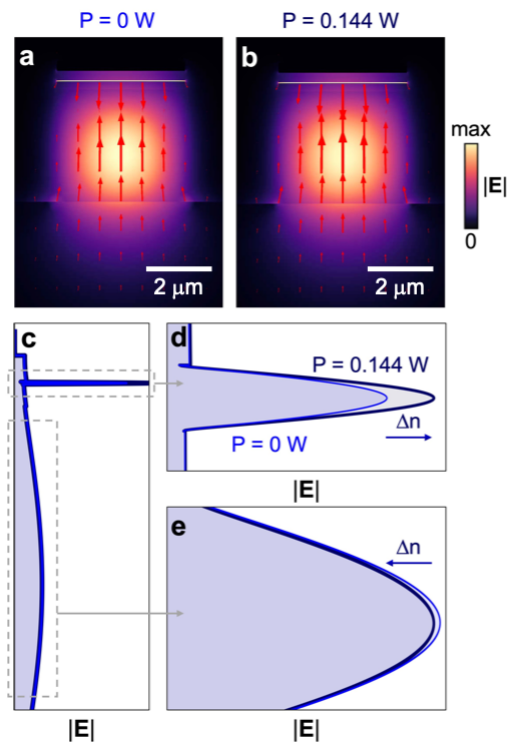}
\caption{\textbf{Linear and nonlinear  eigenmode field distribution with power $P$.} \textbf{(a)} Simulated linear TM eigenmode electric field distribution for $n_0 = 6 \cdot 10^{18}$ cm\textsuperscript{-3} (LBP mode) at $\lambda_0=8$ $\mu$m and a power $P=0$ W. \textbf{(b)} Simulated nonlinear TM eigenmode electric field distribution for $n_0 = 6 \cdot 10^{18}$ cm\textsuperscript{-3} (LBP mode) at $\lambda_0=8$ $\mu$m and a power $P=0.144$ W. \textbf{(c)} Linear and nonlinear eigenmode electric field profiles along the center of the waveguide. \textbf{(d)} Zoomed-in views of the linear and nonlinear eigenmode electric field profiles in the heavily doped InGaAs layer. The mode at $P=0.144$ W shows an appreciable field amplitude increase. \textbf{(e)} Zoomed-in views of the linear and nonlinear eigenmode electric field profiles in the undoped InGaAs region. The mode at $P=0.144$ W shows an appreciable field decrease, associated with a lower mode effective refractive index.} 
\label{fig:figure_eigenmodes}
\end{figure}

The field distributions of the resulting linear and nonlinear eigenmodes for \( n_0 = 6 \times 10^{18} \, \text{cm}^{-3} \), $\lambda_0 = 8$ $\mu$m, and different powers $P$ are shown in Fig. \ref{fig:figure_eigenmodes}. Fig. \ref{fig:figure_eigenmodes}a depicts the linear eigenmode at $P=0$, consistent with the distribution shown in Fig. \ref{fig:1}c. In turn, \ref{fig:figure_eigenmodes}b illustrates the nonlinear eigenmode at \( \mathcal{E}_0 = 1.5 \times 10^5\), corresponding to an input power of \( P = 0.144 \, \text{W} \). The arrow lengths indicate that the LBP's electric field amplitude iincreases with increasing power while the core electric field amplitude decreases. Figure \ref{fig:figure_eigenmodes}c shows the field distribution along a line profile taken through the center of the structure, clearly showing the LBP electric field amplitude in the heavily doped InGaAs layer for the nonlinear mode. This is more clearly seen in the zoomed in field profile shown in Fig. \ref{fig:figure_eigenmodes}d, where a pronounced increase in the electric field is observed in the heavily doped InGaAs layer as the input power increases. In turn, the field distribution of the guided mode in the undoped InGaAs layer shows a decrease of the field amplitude, which is associated to a decrease of the effective mode index in this region. In fact, the overall refractive index change of the nonlinear mode is \( \Delta n = -0.0011 \), that is, the nonlinear response of heavily doped InGaAs causes the refractive index to decrease with increasing power. This behavior is often referred to as self-defocusing, and can stem from various mechanisms, including negative refractive indices in left-handed metamaterials \cite{Hu08_meta}, filamentation and plasma dynamics \cite{Bejot11}, and excitonic interactions in quantum dots \cite{Luo18}, allowing for effects such as the formation of optical vortex solitons \cite{Swartzlander1992} or the design of lenses \cite{Hu08_meta}. Here, it is attributed to free-carrier effects in heavily doped semiconductors. 

By varying $\mathcal{E}_0$ one can get a curve $\Delta n$ as a function of $P$. The resulting values form a linear dependence $\Delta n=n_2(P)P$, where $n_2$ is the nonlinear refractive index, a parameter that quantifies the Kerr nonlinearity of the system. However, \( \Delta n \) does not increase indefinitely with the input power \( P \). In the case of a plane wave propagating through a bulk nonlinear material, the nonlinear effect initially scales linearly with \( P \). The electric field intensity eventually reaches its maximum permissible value before material damage occurs, marking the limit of the nonlinear index change \cite{li2018_nanolasers}. Additionally, our hydrodynamic formalism is derived under the perturbative condition \( n_1 \ll n_0 \). To remain within this regime, we estimate the power at which \( n_1 \) reaches 20\% of \( n_0 \) (see Supplementary Section S2).

\begin{figure}[ht]
\includegraphics[width=0.5\textwidth]{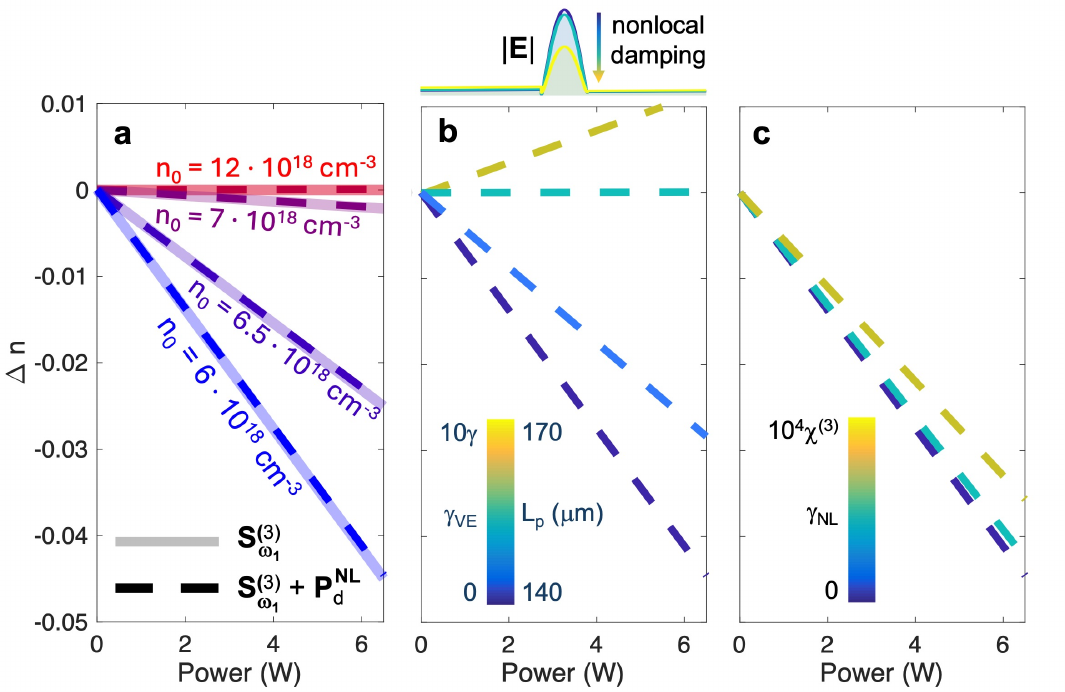}
\caption{\textbf{FE-Kerr-induced change in refractive index $\Delta n$ as a function of input power} $P$. \textbf{(a)} Comparison of refractive index modulation for different carrier concentrations from $n_0 = 6 \times 10^{18}$ (blue) to $n_0 = 12 \times 10^{18} \, \text{cm}^{-3}$ (red). Solid lines correspond to hydrodynamic contributions ($\mathbf{S} ^{(3)}_{\omega _1}$), while dashed lines include lattice nonlinear contributions ($\mathbf{S} ^{(3)}_{\omega _1} + \mathbf{P}^{\mathrm{NL}}_{\mathrm{d}}$). \textbf{(b)} Effect of viscoelastic damping $\gamma_{\text{VE}}$ on the refractive index modulation. As $\gamma_{\text{VE}}$ increases from 0 to $10\gamma$, the field amplitude is reduced, increasing the nonlinear propagation length $L_p$ (color scale). The top panel shows the field distribution $|\textbf{E}|$ as a function of nonlinear damping. \textbf{(c)} Impact of nonlinear damping $\gamma_{\text{NL}}$. The damping parameter $\gamma_{\text{NL}}$ is varied relative to $\chi^{(3)}$ from 0 to $10^4$.} 
\label{fig:fig2}
\end{figure}

The resulting line fits to our calculated $\Delta n (P)$ points are shown in Fig. \ref{fig:fig2}a, comparing the contributions from purely hydrodynamic effects ($\tilde{\mathbf{S}}^{(3)}_{\omega _1}$, solid line) with the combined hydrodynamic and lattice nonlinearity ($\tilde{\mathbf{S}}^{(3)}_{\omega _1} + \mathbf{P}^{\mathrm{NL}}_{\mathrm{d}}$, dashed line) for different doping levels $n_0$. For $n_0 = 12 \times 10^{18} \text{ cm}^{-3}$ (red dashed line), which corresponds to the SPP mode in Fig. \ref{fig:1}d, the modulation due to hydrodynamic contributions remains minimal due to the weak field amplitude inside the active volume (doped InGaAs bulk), as shown in Fig. \ref{fig:1}e. The nonlinear susceptibility $\chi^{(3)}$ contribution remains low in this regime. Conversely, the LBP mode exhibits strong nonlinear modulation, with $\Delta n$ reaching values as high as $-0.04$ for $P = 6.5 \, \text{W}$ at an electron concentration of $n_0 = 6 \times 10^{18} \, \text{cm}^{-3}$ (blue dashed line). This is due to the larger nonlinear active volume and field amplitude of the LBP mode, and corresponds to a relative change of 1.3\%. As anticipated in the description of Fig. \ref{fig:figure_eigenmodes}, the nonlinear response of heavily doped InGaAs causes the refractive index to decrease with increasing power, \textit{i.e.} the slope of $\Delta n$ versus $P$ is negative and thus the Kerr coefficient $n_2$ is negative. Incorporating materials with negative $n_2$ into waveguide structures can enhance performance metrics such as bandwidth, power handling, and signal fidelity. Additionally, electronic contributions to $n_2$ provide a fast, instantaneous negative nonlinear response, which is highly advantageous for ultrafast optical applications. Moreover, the presence of a positive $\chi^{(3)}$ component introduces a slight upward bending of the line, but it is orders of magnitude weaker than the hydrodynamic contribution. Finally, moving away from the LBP resonance at $\lambda_0 = 8$ $\mu$m, the hydrodynamic Kerr nonlinearity is already much weaker, as shown by calculations at $n_0 = 6.5$ and 7 $\times 10^{18} \text{ cm}^{-3}$. For \(n_0 = 5 \times 10^{18} \, \text{cm}^{-3}\), the values decrease and are hidden behind those for \(n_0 = 12 \times 10^{18} \, \text{cm}^{-3}\) in the plot. This suggests that the large nonlinearity at \(n_0 = 6 \times 10^{18} \, \text{cm}^{-3}\) is not solely due to the \(n_0^{-2}\) dependence of the FE Kerr effect but is mainly caused by the resonance of the LBP.

We now study the robustness of the $\Delta n$ driven by the LBP mode under more realistic conditions. It has been reported that materials like CdO and ITO, particularly near their ENZ frequencies, often exhibit viscosity-driven nonlocal effects in their optical response \cite{de2018viscoelastic}. To account for this effect, we introduce a nonlocal viscoelastic damping coefficient $\beta = \sqrt{\frac{3}{5}-i\frac{\gamma_\text{VE}}{\omega}\frac{4}{15}}v_F$, which accounts for both dispersive and dissipative contributions \cite{de2018viscoelastic}. Figure \ref{fig:fig2}b depicts the impact of viscoelastic damping on the nonlinear modulation for increasing values of $\gamma_{\text{VE}}$ relative to the Drude damping rate $\gamma$ of heavily doped InGaAs (color gradient) ranging from $0$ to $10 \gamma$. Here we fix a doping level \(n_0 = 6 \times 10^{18} \, \text{cm}^{-3}\), which corresponds to the maximum FE Kerr nonlinearity in Fig. \ref{fig:fig2}a. As $\gamma_{\text{VE}}$ increases, the field amplitude in the heavily doped InGaAs layer diminishes, resulting in a reduced nonlinear index shift and an increase in the nonlinear propagation length $L_p$, from $L_p=141$ $\mu$m for $\gamma_{\text{VE}}=0$ up to $L_p=170$ $\mu$m $\gamma_{\text{VE}}=10\gamma$ (color bar). Despite this reduction, the overall nonlinearity remains robust. Another source of damping is a so-called nonlinear damping, which can be modeled by an additional imaginary term $i \gamma_{\text{NL}} |\textbf{E}|^2\textbf{E}$, giving rise to a change in the refractive index $\Delta n + \Delta n_{\gamma_\text{NL}}$. Figure \ref{fig:fig2}c shows the influence of $\gamma_{\text{NL}}$ relative to $\chi^{(3)}$, ranging from $0$ to $10^4\chi^{(3)}$. Again, we fix a doping level \(n_0 = 6 \times 10^{18} \, \text{cm}^{-3}\), \textit{i.e.}, the doping yielding the maximum FE Kerr nonlinearity in Fig. \ref{fig:fig2}a. We observe that the change in the refractive index $\Delta n + \Delta n_{\gamma_\text{NL}}$ decreases for increasing values of $\gamma_\text{NL}$. We note that $\gamma_\text{NL}$ can induce a red shift of the LBP resonance due to heating of electrons in the conduction band. In addition, since $\chi^{(3)}$ is considerably smaller than the hydrodynamic contribution, the nonlinear damping has a small impact on the refractive index modulation, confirming the resilience of the optical nonlinearity in this system.

Finally, we compute the waveguide’s Kerr nonlinear coefficient \(\gamma_{\mathrm{vg}}\), which quantifies the waveguide’s nonlinear response to an optical field and is given by \cite{Li2016_figure,Afshar09}:
\begin{equation}
    \gamma_\mathrm{wg} = k_1' 
     \frac{\varepsilon_0}{\mu_0}
    \frac{\int n^2(\textbf{x}, \textbf{y})n_2^2(\textbf{x}, \textbf{y}) \left( 2|\textbf{E}_\nu|^4 + |\textbf{E}_\nu^2|^2 \right) \, dA}{
    3 \left( \int (\textbf{E}_\nu \times \textbf{H}_\nu^*) \cdot \hat{\mathbf{z}} \, dA \right)^2},
    \label{eq:NLcoef_wg}
\end{equation}
where \(k_1'\) is given by Eq. \eqref{eq:nonlinear_eigenvalue}, \( k_0 \) is the free-space wavevector, \(n(\mathbf{x}, \mathbf{y})\) is the refractive index profile, and \(\mathbf{E}_\nu\) and \(\mathbf{H}_\nu\) are the electric and magnetic fields of mode with index $\nu$, respectively. The integral in the numerator captures the spatial overlap of the nonlinear interaction with the mode intensity, whereas the denominator normalizes by the mode power. The mode volume is defined as $V = \frac{\int \mathbf{S} \cdot \hat{\mathbf{z}} \, dA}{\max(\varepsilon|\mathbf{E}|^2)}$, where \(\mathbf{S} = \mathbf{E} \times \mathbf{H}^*\) is the Poynting vector. For a carrier concentration of \(n_0 = 6 \times 10^{18}\,\mathrm{cm}^{-3}\) and a wavelength of \(\lambda_0 = 8\,\mu\mathrm{m}\) (corresponding to the LBP mode in Fig.\,\ref{fig:1}c), the mode volume is \(V = 0.47\,\mu\mathrm{m}^2\). This exceptionally small mode volume arises from the strong field confinement in the heavily doped InGaAs layer (Fig.\,\ref{fig:1}e). Consequently, we find \(\gamma_{\mathrm{vg}} \approx 4 \times 10^7\,\mathrm{W}^{-1}\,\mathrm{km}^{-1}\), positioning LBP-based plasmonic waveguides as a promising platform for nonlinear integrated photonic circuits (see Table \ref{tab:nonlinearcoeffs}). For comparison, silicon and AlGaAs waveguides at 1.55\,\(\mu\mathrm{m}\) typically exhibit nonlinear parameters of \(10^3\text{--}10^5\,\mathrm{W}^{-1}\,\mathrm{km}^{-1}\), while silicon nitride and lithium niobate show more moderate values of \(10^0\text{--}10^2\,\mathrm{W}^{-1}\,\mathrm{km}^{-1}\). On the other hand, ITO waveguides can reach values of up to 10\textsuperscript{7} by leveraging ENZ resonances \cite{Li20_ENZ_ITO,RojasYanez25}. Our significantly higher value at 8 $\mu$m arises from the intrinsically large nonlinear refractive index in the mid-infrared and the pronounced field confinement that reduces the effective mode area. This synergy of high nonlinearity and tight mode confinement drives $\gamma_\mathrm{wg}$ far beyond near-infrared benchmarks, underscoring the potential of FE nonlinearities in heavily doped semiconductors for mid-IR integrated photonic platforms. The problem of free carrier losses, typically increasing as $\lambda^{-2}$, becoming more severe at mid-IR wavelengths, can be mitigated with hybrid designs featuring intrinsic cores and evanescently coupled doped layers, like the one presented in this work.

\begin{table}[ht]
\centering
\caption{State-of-the-art nonlinear coefficients $\gamma_\mathrm{wg}$ 
         for various waveguide platforms.}
\begin{tabular}{lccc}
\hline
waveguide platform & $\lambda_0$ ($\mu$m) & $\gamma_\mathrm{wg}$ (W$^{-1}$\,km$^{-1}$) \\
\hline
Si                               & 1.55           & $10^3 - 10^5$              \\
SiN                              & 1.55           & $10^1 - 10^2$              \\
AlGaAs                           & 1.55           & $10^3 - 10^5$              \\
LiNbO$_3$                        & 1.55           & $10^0 - 10^1$              \\
Si–organic slot waveguides with\\
\quad nonlinear polymers         & 1.55           & $7 \times 10^6$            \\
Nanoshell plasmonic waveguides   & 1.55           & $4.1 \times 10^4$          \\
Metal-ITO-metal slot waveguides  & $\approx$ 1.4  & $10^7$                     \\ 
\textbf{This work}               & \textbf{8}     & $\mathbf{4 \times 10^7}$   \\
\hline
\end{tabular}
\label{tab:nonlinearcoeffs}
\end{table}

Finally, we calculate the FOM of the conversion efficiency $\eta$ at a length $L$, driven by a power of $P$, which can be calculated as $\eta \approx e^{-\alpha L} \left( \gamma_\mathrm{wg} P L_{\text{eff}} \right)^2$ \cite{Li2016_figure,Li2018_fundamentals}, where the attenuation and nonlinear coefficients, $\alpha$ and $\gamma_\mathrm{wg}$ respectively, are calculated from the electromagnetic field using mode analysis. $\alpha$ is given by $1/L_p$, where $L_p$ for the LBP is 140 $\mu$m, and $\gamma_\mathrm{wg}$ is given by Eq. \eqref{eq:NLcoef_wg}. The effective length $L_{\text{eff}}$ is defined as $L_{\text{eff}} = \frac{1 - e^{-\alpha L}}{\alpha}$, in agreement with the one reported for lossy waveguides \cite{Li2016_figure}. The conversion efficiency $\eta$ takes into account the balance between nonlinearity and losses. For instance, silicon and silicon nitride waveguides typically yield values of $\eta \approx$ 10\textsuperscript{-2} to 1. However, silicon tends to attain higher $\eta$ at short lengths due to its strong nonlinearity, but losses quickly reduce efficiency for longer waveguides. Silicon nitride waveguides, in turn, with lower $\gamma_\text{wg}$, compensate with much lower loss and longer effective interaction lengths, allowing significant nonlinear effects over longer distances. On the other hand, metal-ITO-metal slot waveguides, with a very high $\gamma\approx$ $10^7$ W$^{-1}$\,km$^{-1}$ yield relatively low $\eta$ values due to its large losses, which limit propagation length to a few microns, as well as phase accumulation \cite{Li20_ENZ_ITO,RojasYanez25}. For our system, the maximum $\eta\approx 200$ is attained at a waveguide length $L\approx 160$ $\mu$m, \textit{i.e.}, at a very long propagation length, for a power $P=6.5$ W (see Supplementary section S3). Values of $\eta$ around 100 to 200 are comparable to or exceed state-of-the-art values at telecom wavelengths, but at mid-IR, where losses tend to be higher, and also exceed Kerr conversion efficiencies reported for plasmonic waveguides by several orders of magnitude \cite{Li2016_figure,Li2018_fundamentals}. This suggests efficient nonlinear interactions by exploiting the low-loss character of our all-semiconductor waveguides, together with strong FE nonlinearities from heavily doped InGaAs.

\begin{figure*}[!ht]
\includegraphics[width=0.85\textwidth]{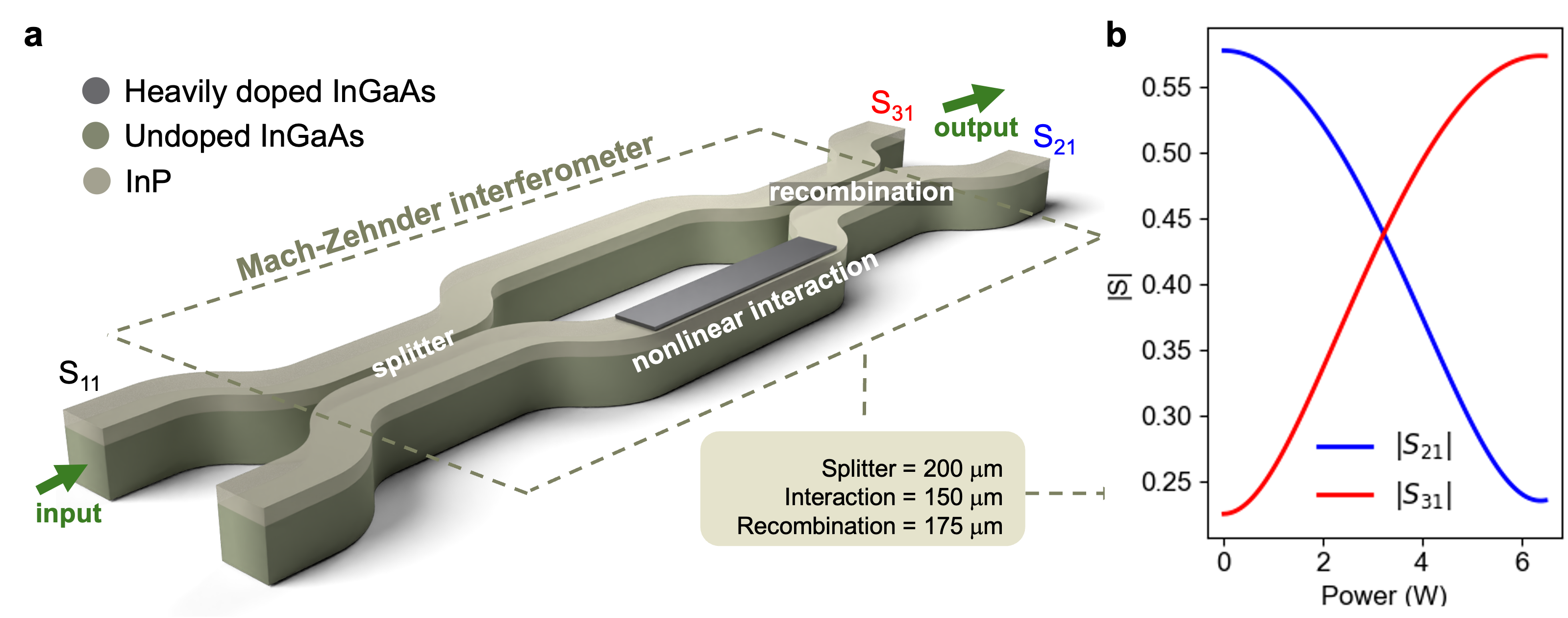}
\caption{\textbf{FE-Kerr-based Mach-Zehnder interferometer design and power-dependent transmittance modulation.}  
\textbf{(a)} 3D schematic of the Mach-Zehnder interferometer (MZI) composed of undoped InGaAs (green) cladded by InP (beige). Input light enters through port \( S_{11} \) (green arrow) and is split into two arms at the splitter (200 $\mu$m length). The right arm includes a 30-nm-thick heavily doped InGaAs layer (dark gray) to induce an intensity-dependent refractive index change, while the left arm does not. The nonlinear interaction region (150 $\mu$m) in the doped arm exploits hydrodynamic and Kerr nonlinearities to produce a phase shift, while the left arm undergoes only dielectric Kerr effects. The two arms interfere at the output in a 150 $\mu$m-long recombination section, determining the transmittance at output ports \( S_{21} \) (blue) and \( S_{31} \) (red).  
\textbf{(b)} Power-dependent output transmittance, showing the modulation at ports \( S_{21} \) and \( S_{31} \) as a function of input power.}
\label{fig:fig3}
\end{figure*}

\section{FE-Kerr--induced TRANSMITTANCE MODULATION in A hybrid heavily doped semiconductor MACH-ZEHNDER INTERFEROMETER}

Finally, we investigate the possibility of leveraging the nonlinear modulation of the refractive index discussed in Fig. \ref{fig:fig2} to induce an intensity-dependance modulation of the transmittance spectrum. To do so, we employ a Mach-Zehnder interferometer (MZI), sketched in Fig. \ref{fig:fig3}a. The input light (green arrow) enters through port \( S_{11} \) into the waveguide core, made of undoped InGaAs (green) cladded by InP (beige), with a mode index $n_{\text{eff}}=3.11$ (see Supplementary section S1) and propagation losses $\gamma/20$, where \( \gamma\) corresponds to the Drude damping for heavily doped InGaAs, leading to a propagation length $L_p=32$. mm. The input light is then divided at a 250 $\mu$m-long splitter, with a curvature radius $R=400$ $\mu$m optimized to minimize reflection and scattering losses. The two waveguide arms guide the light along different paths: the right arm contains a 30-nm-thick layer of heavily doped InGaAs on top (gray), inducing a refractive index change dependent on the input intensity based on the hydrodynamic Kerr plus the third-order susceptibility \( \chi^{(3)} \), while the left arm remains undoped and undergoes only Kerr effects characterized by the third-order susceptibility $\chi^{(3)}$. Therefore, the key contribution to the phase shift difference comes from the hydrodynamic nonlinearity in the right arm. Traversing a 175 $\mu$m-long recombination section, the two arms interfere at the output ports \( S_{21} \) (blue path) and \( S_{31} \) (red path), with the phase accumulation in the left arm given by \( \phi_{\text{l}} = k_0 n_{\text{eff}} L \), where \( L \) is the length of the nonlinear interaction section, set to 150 $\mu$m. For the arm with the heavily doped InGaAs, the phase accumulation depends on the input intensity as $\phi_{\text{r}} = \int_{0}^{L} k_0 \Delta n(z) \, dz = \int_{0}^{L} k_0 n_2 P(z) \, dz$, where $n_2=-0.007$ W\textsuperscript{-1} is the nonlinear refractive index coefficient extracted from the slope of the blue curve in Fig. \ref{fig:fig2}a, and $P(z)$ is the power profile along the waveguide. Since the electric field intensity \( E_0^2 \) decays due to losses, the power profile is modeled as \( P(z) = P_0 \exp(-\gamma z) \), where \( P_0 \) is the input power and $\gamma=\text{Im}(n_\text{eff})k_0$ is the attenuation constant of the mode, with $\text{Im}(n_\text{eff})=0.009$. Substituting this into the integral yields $\phi_{\text{r}} = \frac{k_0 n_2 P_0}{\gamma} \left[1 - \exp(-\gamma L) \right]$. The relative phase difference between the two arms is \( \Delta \phi = \phi_{\text{r}} - \phi_{\text{l}} \), which dictates the interference pattern at the output. Fig. \ref{fig:fig3}b shows the power-dependent transmittance at the output ports: the blue curve $|S_{21}|$ represents the transmittance through port 2, while the red curve $|S_{31}|$ represents the transmittance through port 3. At low input powers, nonlinearities are low, and the transmittance is dominated by the linear arm, with most light exiting through port 2, $|S_{21}| \approx 60\%$. As input power increases, the nonlinear phase shift in the doped arm grows, shifting the interference toward destructive interference at port 2 and constructive interference at port 3. At around 3 W input power, the output switches entirely, with most light starting to exit through port 3, up to $|S_{31}| \approx 60\%$. The modulation range from $\approx 20\%$ to $\approx 60\%$ demonstrates the significant impact of the hydrodynamic Kerr effect on the transmittance. Importantly, this transmittance could be potentially tuned on-demand by leveraging the reconfigurability of the FE Kerr effect through the application of a bias voltage \cite{hu2024modulating}.

\section{Conclusions}

In conclusion, our results demonstrate that FEs can induce exceptionally strong Kerr nonlinearities in all-semiconductor hybrid waveguides. Building on this, we can design waveguides that harness LBPs, nonlocal resonances, that achieve ultrahigh nonlinear coefficients. Leveraging this high nonlinearity, we have shown an all-optical intensity-dependent modulation of the transmittance in a MZI. These results highlight the potential of FE nonlinearities in heavily doped semiconductors for integrated nonlinear photonics, offering a new route to realize high-speed, low-power, and ultra-compact optical devices for next-generation data processing.

\vspace{0.6cm}

\section*{Acknowledgements}

We acknowledge funding from the European Innovation Council through its Horizon Europe program with Grant Agreement No. 101046329. Views and opinions expressed are those of the authors only and do not necessarily reflect those of the European Union. Neither the European Union nor the granting authority can be held responsible for them.

\bibliography{apssamp.bib}

\end{document}